\documentclass[prb, twocolumn, showpacs, footinbib]{revtex4-1}
\usepackage{float}
\usepackage{bbm}
\usepackage{graphicx}
\usepackage{subcaption}
\usepackage{bm}
\usepackage{amsmath,amssymb}
\usepackage{color}
\usepackage{hyperref}
\usepackage{relsize}
\usepackage{mathtools}

\DeclarePairedDelimiter\ket{\lvert}{\rangle}
\DeclarePairedDelimiterX\braket[2]{\langle}{\rangle}{#1 \delimsize\vert #2}
\begin{document}

\title{A robust operating point for capacitively coupled singlet-triplet qubits}

\author{M.~A.~Wolfe$^1$}
\author{F.~A.~Calderon-Vargas$^2$}
\author{J.~P.~Kestner$^1$}
\email{jkestner@umbc.edu}
\affiliation{$^1$Department of Physics, University of Maryland Baltimore County, Baltimore, MD 21250, USA\\
$^2$ Department of Physics, Virginia Tech, Blacksburg, VA 24061, USA}

\begin{abstract}
Singlet-triplet qubits in lateral quantum dots in semiconductor heterostructures exhibit high-fidelity single-qubit gates via exchange interactions and magnetic field gradients. High-fidelity two-qubit entangling gates are challenging to generate since weak interqubit interactions result in slow gates that accumulate error in the presence of noise. However, the interqubit electrostatic interaction also produces a shift in the local double well detunings, effectively changing the dependence of exchange on the gate voltages.  We consider an operating point where the effective exchange is first order insensitive to charge fluctuations while maintaining nonzero interactions.  This ``sweet spot" exists only in the presence of interactions.  We show that working at the interacting sweet spot can directly produce maximally entangling gates and we simulate the gate evolution under realistic $1/f$ noise. We report theoretical two-qubit gate fidelities above 99\% in GaAs and Si systems.
\end{abstract}

\maketitle

\section{Introduction}
The singlet-triplet qubit \cite{Petta2005} is an attractive platform for quantum information processing due to its fast single-qubit operations \cite{Foletti2009} and extended coherence times \cite{Bluhm2010,Maune2012}. Voltage gates ``detune" the double quantum dot (DQD), i.e., adjust the energy difference $\varepsilon$ between the two minima of the DQD potential, driving rotations around the $z$ axis of the Bloch sphere via the exchange interaction $J(\varepsilon)$, and rotations about the $x$ axis are induced by a magnetic field difference across the DQD, $h$. Two-qubit entangling operations can be achieved via capacitive coupling \cite{Stepanenko2007} or interqubit exchange interaction \cite{Wardrop2014}. Here, we consider the former because they have been demonstrated experimentally\cite{Shulman2012,Nichol2017} and are naturally robust to leakage outside of the logical subspace. The primary source of error is fluctuation of the detuning due to charge noise in the device, inhibiting the singlet-triplet qubit from performing at fault tolerant levels.
	
Single-qubit gates can be fast, with a $\pi$-rotation about the $z$ axis demonstrated in 350 ps \cite{Petta2005}, and precise, with 99\% fidelity \cite{Nichol2017}. However, the relatively weak capacitive interaction generates two-qubit gates that are much slower, 140 ns for a controlled-phase ({\sc cphase}) gate\cite{Shulman2012}, and these slow gates accumulate substantial errors in the presence of noise, with fidelities yet to exceed 90\% \cite{Nichol2017}. Strategies such as dynamical decoupling\cite{Dial2013}, pulse shaping \cite{Barnes2015,Zeng2017}, composite pulse sequences \cite{Sunny2012,Kestner2013,CalderonVargasPRL2017}, and control tuning using iterative experimental feedback \cite{Cerfontaine2016} can improve the fidelity of gating in the presence of noise. These methods are particularly effective against noise that fluctuates on timescales much longer than the time required to complete the quantum operation (i.e., the gate time). For the slower two-qubit gates, however, high-frequency charge noise is difficult to suppress. An alternative approach is to use a robust operating point in control parameter space, often called a ``sweet spot," where the Hamiltonian is insensitive to certain perturbations and hence the effect of fluctuations of any frequency is reduced.  The remarkable recent progress of superconducting qubits can be largely attributed to the introduction of the transmon sweet spot \cite{Koch2007}.  In this Rapid Communication we introduce such a sweet spot for two coupled singlet-triplet qubits.
	
Previous investigations of sweet spots in singlet-triplet qubits have mainly focused on a single, isolated qubit \cite{Stopa2008, Li2010, Nielsen2010, Koh2013, Wong2015, Hiltunen2015, Reed2016, Martins2016}, in which case the sweet spot previously discussed is not appropriate for capacitive coupling and the sweet spot we present does not exist. Where the case of interacting qubits has been considered \cite{Yang2011,Nielsen2012}, the focus was on the robustness of the coupling term in the Hamiltonian. However, the primary contribution to the error during a two-qubit gate is from fluctuations of the strong local terms in the Hamiltonian rather than in the weak coupling term itself.
	
In the present work, using harmonic oscillator basis functions in a Hund-Mulliken (HM) model for the singlet-triplet two-qubit Hamiltonian \cite{Ramon2011,Fernando2015}, we report an interacting sweet spot where the local effective exchange terms are insensitive to fluctuations in the detunings caused by charge noise, while maintaining a nonzero two-qubit coupling. Our results are meant to be taken qualitatively, as computational methods such as exact diagonalization\cite{Hiltunen2014} or a full configuration interaction\cite{Nielsen2012} would be necessary for quantitative precision. However, since the experimental potential profile is typically not known precisely anyway, a qualitative approach is not inappropriate and provides a starting point for experimental fine tuning. While the sweet spot only suppresses charge noise, it can be combined with standard echo pulses to also mitigate magnetic field gradient noise, thus producing high-fidelity two-qubit entangling gates. We perform numerical simulations of performance at the interacting sweet spot in the presence of realistic noise with parameters typical for GaAs and Si devices, and find that fidelities above 99\% are achievable simply by choosing the operating parameters wisely.

\section{Sweet Spot Analysis}
\begin{figure}
	\includegraphics[width = .9\columnwidth]{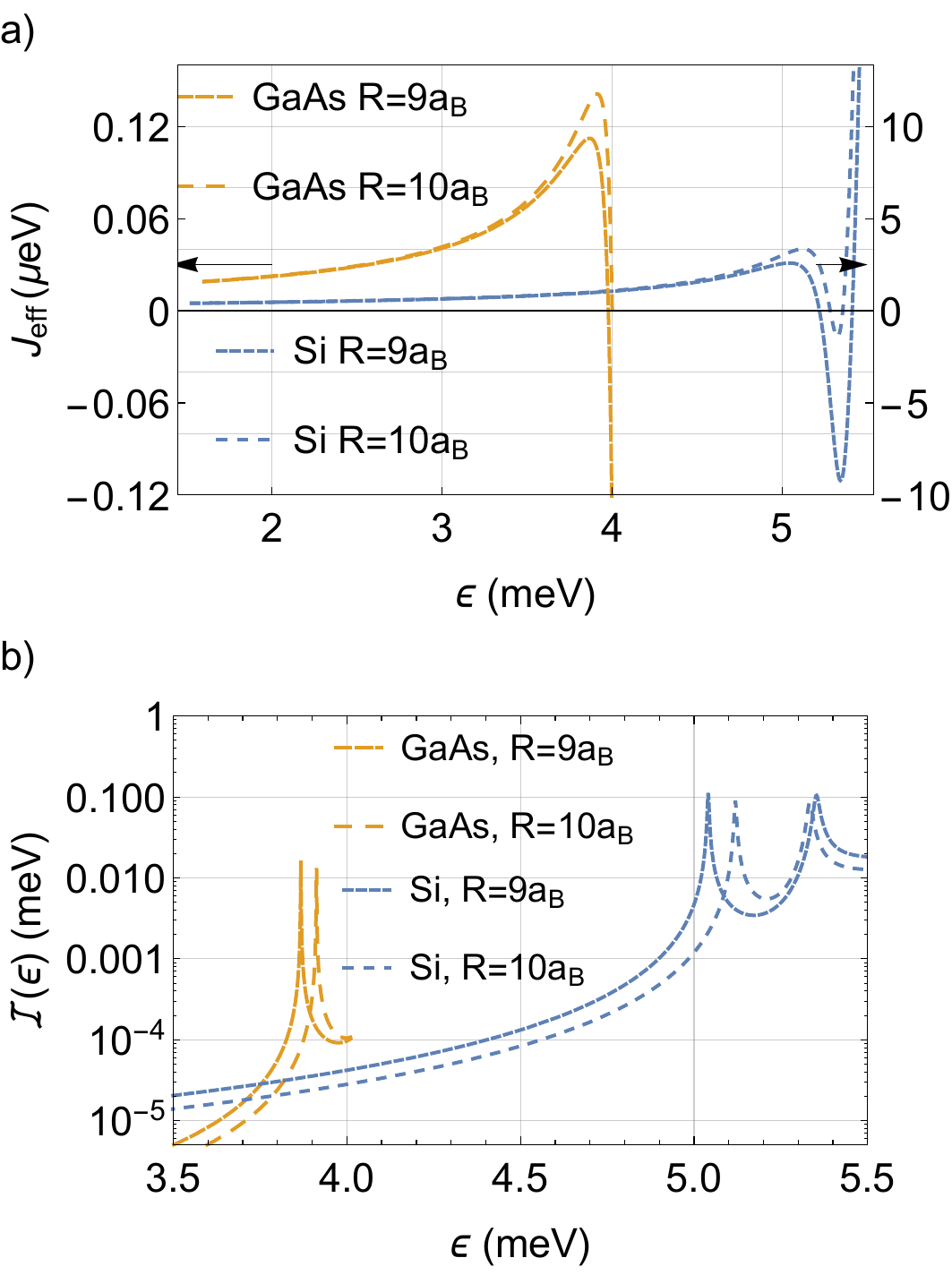}
	\caption{(Color online.) \textbf{(a)} Two examples of effective exchange vs symmetric detuning $\varepsilon_1=\varepsilon_2=\varepsilon$ for GaAs (orange) and Si (blue) at two interqubit separations (solid/dashed). The interacting sweet spots are located extremely close to the zeros of $J_{eff}'(\varepsilon)$. In the Si case, we focus on the first extremum since the HM model is more accurate at lower detunings. \textbf{(b)} Insensitivity defined in Eq.~\eqref{insensitivity} as a function of symmetric detuning on the $\varepsilon_1$--$\varepsilon_2$ diagonal line for two examples of interqubit separation.}\label{fig:fig1}
\end{figure}

The Hamiltonian for two capacitively coupled singlet-triplet qubits in a linear four-dot array is written using a HM approximation\cite{Stepanenko2007,Ramon2011,Fernando2015}, where the two-qubit Hilbert space is spanned by products of the unpolarized triplet state $\ket{T_0}$ and the hybridized singlet state $\ket{\tilde{S}}$ formed by the lowest harmonic oscillator orbitals centered at each minima, where the singlet contains a small admixture of a doubly occupied orbital controlled by the detuning of the corresponding DQD, $\varepsilon_i$. In the basis $\left\{\ket{\tilde{S}\tilde{S}},\ket{\tilde{S}T_0},\ket{T_0\tilde{S}},\ket{T_0T_0}\right\}$,
\begin{multline}\label{Two-qubit_Hamiltonian}
H(\varepsilon_1,\varepsilon_2,h_1,h_2) = \sum_{i=1}^2 \biggl[\left(\frac{J_i(\varepsilon_i)}{2}-\beta_i(\varepsilon_1,\varepsilon_2)\right)\sigma_z^{(i)} \\
+ \frac{h_i}{2}\sigma_x^{(i)}\biggr]
+ \alpha\left(\varepsilon_1,\varepsilon_2\right)\sigma_z^{(1)}\sigma_z^{(2)},
\end{multline}
where $\vec{\sigma}^{(i)}$ are the Pauli operators for qubit $i$, $J_i$ is the local exchange splitting, and the interqubit Coulomb interactions contribute both a local shift in the effective exchange due to a monopole-dipole interaction, $\beta_i(\varepsilon_1,\varepsilon_2)$, and a non-local term due to a dipole-dipole interaction $\alpha(\varepsilon_1,\varepsilon_2)$. The magnetic field difference between the two wells of each DQD is $h_i$, which, in GaAs, originates from an inhomogeneous Overhauser field, $h_i=h\approx 2\pi \times$ 30 MHz\cite{Shulman2012}, while in Si comparable values can be realized with integrated micromagnets,  $h_i=h\approx 2\pi \times$ 15 MHz\cite{Eriksson2014}. We use the convention that $\varepsilon = 0$ corresponds to a symmetric double well and $\varepsilon > 0$ raises (lowers) the inner (outer) of the two dots. Pulsing both qubits to positive $\varepsilon$ then corresponds to tilting the DQDs away from each other. Charge noise enters the Hamiltonian through $J$, $\beta$, and $\alpha$ terms since they are controlled electrically via $\varepsilon_1$ and $\varepsilon_2$. It has been found empirically in single-qubit experiments that the exchange splitting increases roughly exponentially with detuning, $J_i(\varepsilon_i) \propto J'_i(\varepsilon_i)$ \cite{Dial2013}. Thus, while large detuning generates fast gates, the sensitivity to charge noise, proportional to $J'_i(\varepsilon_i)$, also increases with detuning. However, when the Coulomb interaction from the neighboring qubit is considered, the effective exchange for qubit $i$, $J_{\mathrm{eff},i}(\varepsilon_1,\varepsilon_2) \equiv J_i(\varepsilon_i)/2 - \beta_i(\varepsilon_1,\varepsilon_2)$, can markedly deviate from simple exponential behavior due to the monopole-dipole interaction \cite{Ramon2011,Fernando2015}, as shown in Fig.~\ref{fig:fig1}a.

To model experiments in GaAs\cite{Shulman2012} we take relative permittivity $\kappa$ = 13.1$\epsilon_0$, effective electron mass $m^* = 0.067m_e$, confinement energy of the quantum dots $\hbar\omega_0 = $ 1meV (hence, an effective Bohr radius $a_B=\sqrt{\hbar/m^*\omega_0}\approx$ 34nm), and intraqubit distance $2a=5a_B$ ($a$ denoting the distance from center to minima of the DQD in the (1,1) charge configuration). Appropriately sized harmonic oscillator orbitals give an on-site interaction energy $U=4.1$meV.  The tunneling rate, $t_0$, is computed assuming a quartic DQD potential \cite{Fernando2015}, resulting in $t_0=7\mu$eV. The resulting exchange model is near the limit of the validity of the HM approximation, as can be tested by checking the monotonicity of exchange with intraqubit distance \cite{Li2010}, but the qualitative physics is still safely captured.

In Si, recent overlapping Al gate layering techniques have produced more densely packed dots with size $a_B = $ 29nm and spacing $2a\simeq 3.5 a_B$ \cite{Petta2015}.  Using these parameters, $\kappa = 11.68\epsilon_0$, and $m^* = 0.19m_e$, one estimates $U= 5.3$meV.  Again assuming a quartic DQD potential would produce a tunneling rate outside the validity of HM, but since the tunneling barrier can be independently tuned \cite{Simmons2009,Reed2016}, it is more reasonable anyways to directly set the tunneling parameter to an experimentally reported value. We take $t_0$ = 40$\mu$eV \cite{Simmons2009}, which allows a good approximation for an intraqubit distance of $2a$ = 160nm.

The sensitivity of the Hamiltonian to charge noise is quantified by the Frobenius norm of the gradient on the $\varepsilon_1$--$\varepsilon_2$ plane,
\begin{equation}\label{sensitivity}
\left\|\vec{\nabla} H(\varepsilon_1,\varepsilon_2)\right\| \simeq \sqrt{
	\mathlarger{\sum}_{\{i,j\}=1}^2 \left(\frac{\partial J_{\mathrm{eff,i}}}{\partial\varepsilon_j}\right)^2},
\end{equation}
where we have omitted the derivatives of $\alpha$ because they are orders of magnitude smaller, as also noted experimentally\cite{Shulman2012}, and their effect is negligible. (This makes Eq.~\eqref{sensitivity} easier to measure experimentally too.) Numerically minimizing this function using the HM form of $J_{\mathrm{eff,i}}$ derived in Ref.~\onlinecite{Fernando2015}, we find previously unreported minima that reduce the sensitivity by orders of magnitude.  The locations of these interacting sweet spots depend on the interqubit distance $2R$ (i.e., the center-to-center distance between the DQDs in the (1,1,1,1) charge configuration), but always lie on the $\varepsilon_1=\varepsilon_2$ diagonal line in our search space. Therefore, from here on, we only consider symmetric operating points $\varepsilon_1=\varepsilon_2=\varepsilon$ (though the fluctuations in $\varepsilon_i$ are \emph{not} restricted to be symmetric). At the sweet spot, $\varepsilon_{ss}$, $J_{\mathrm{eff}}'\left(\varepsilon_{ss}\right) \approx 0$ while $\alpha\left(\varepsilon_{ss}\right) \neq 0$ \footnote{Note that Ref.~\onlinecite{Ramon2011} also discusses a sweet spot resulting from the \protect{$\beta$} terms, but it is where \protect{$J_{\text{eff}}=0$} rather than where its derivative is zero, so it is not the same in that it is not a robust operating point.}. This only occurs when interqubit interactions are present and, if one stays predominately in the (1,1,1,1) configuration, only when the qubits are tilted away from each other (i.e., in a ``breathing" mode rather than a ``sloshing" mode). Other sweet spots may exist at larger, possibly asymmetric, detunings, but we cannot use HM to explore those regions.

In order to capture the effect of charge noise on an entangling gate, we define the two-qubit insensitivity in analogy to the single-qubit case of Ref.~\onlinecite{Reed2016} as something roughly akin to the rate of entanglement divided by the rate of decoherence,
\begin{equation}\label{insensitivity}
\mathcal{I}(\varepsilon) = \frac{\alpha(\varepsilon)}{\left\|\vec{\nabla} H(\varepsilon)\right\|}.
\end{equation}
Fig.~\ref{fig:fig1}b shows that the insensitivity, though finite, increases by orders of magnitude at an interacting sweet spot.

\begin{figure}
	\includegraphics[width=\columnwidth]{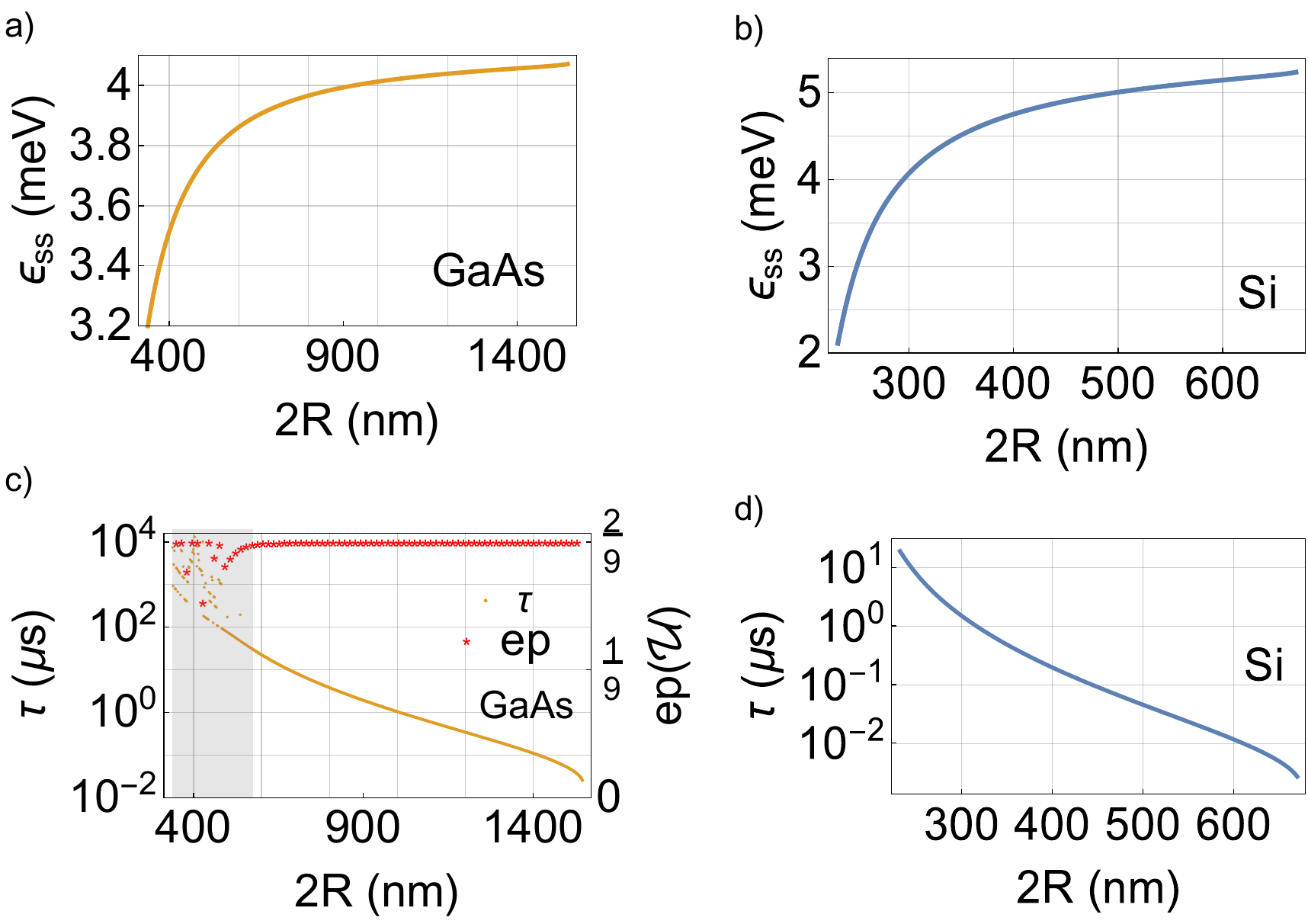}
	\caption{(Color online.) Interacting sweet spot operating point vs interqubit distance $2R$ for \textbf{(a)} GaAs and \textbf{(b)} Si. Log plot of maximally entangling gate time vs $2R$ for GaAs \textbf{(c)} and Si \textbf{(d)}. For GaAs, the gray region is where the entanglement power (red) could not consistently be optimized to within 1\% of the maximum value 2/9.}\label{fig:fig2}
\end{figure}

The HM approximation can break down at high detuning if the S(0,2) probability is large, so we restrict detunings such that this probability is below $\hbar\omega_0/U$ (which is why the GaAs curves in Fig.~\ref{fig:fig1} appear truncated). For Si, two sweet spots are valid within this range, however, we focus on the lower one because it is deeper in the regime of HM applicability.

The location of the interacting sweet spot depends on the interqubit separation, $2R$, as shown in Figs.~\ref{fig:fig2}a and \ref{fig:fig2}b. For the parameters given above, we find no interacting sweet spots at interqubit distances greater than $2R_{\text{max}}$ = 1544nm (674nm) for GaAs (Si). We do not consider interqubit distances less than four times the intraqubit distance, $2R_{\mathrm{min}}=4a$, so as to safely neglect tunneling between adjacent DQDs.

Operating at these new sweet spots is useful because it generates an entangling operation while providing protection against charge noise. The entangling power can be quantified as\cite{Balakrishnan2010}
\begin{equation}\label{ep}
  \text{ep}(U) = \frac{2}{9}\left(1-\left|\frac{\mathrm{tr}^2[(Q^{\dagger}UQ)^{\mathrm{T}}Q^{\dagger}UQ]}{16}\right|\right),
\end{equation}
where $Q$ is the transformation from the logical basis to the Bell basis\cite{Makhlin2002}. One can always construct a controlled-NOT ({\sc cnot}) gate from at most two applications of any maximally entangling gate \cite{Williams2010}. We numerically search for the shortest time $\tau$ required to generate a maximally entangling gate in a single square pulse of the detunings to the sweet spot, $\varepsilon_{ss}$.  The results are shown in Figs.~\ref{fig:fig2}c and \ref{fig:fig2}d. For our GaAs parameters, there is a region of interqubit distances where no sweet spot gate can directly generate more than 99\% of the maximal entangling power. For our Si parameters, there is no such excluded region, and gate times are generally faster in Si due to the smaller distance scale. Note that gate time actually decreases as the qubits are moved farther apart. Although this may seem counterintuitive at first, it can be understood by noting that increasing the interqubit distance moves the interacting sweet spot to stronger detuning.  Since $\tau \propto 1/\alpha(\varepsilon)$, and the nonlocal coupling $\alpha(\varepsilon)$ increases exponentially with the detuning (since $\alpha(\varepsilon) \propto J_1(\varepsilon_1)J_2(\varepsilon_2)$ has been shown empirically\cite{Shulman2012}) but, as an electrostatic term, only decreases polynomially with distance, if we restrict ourselves to operations \emph{at the sweet spot}, the gate time will decrease as the interqubit distance increases.

\section{Simulations}\label{simulation}
\begin{figure}
	\centering
	\includegraphics[width = .95\columnwidth]{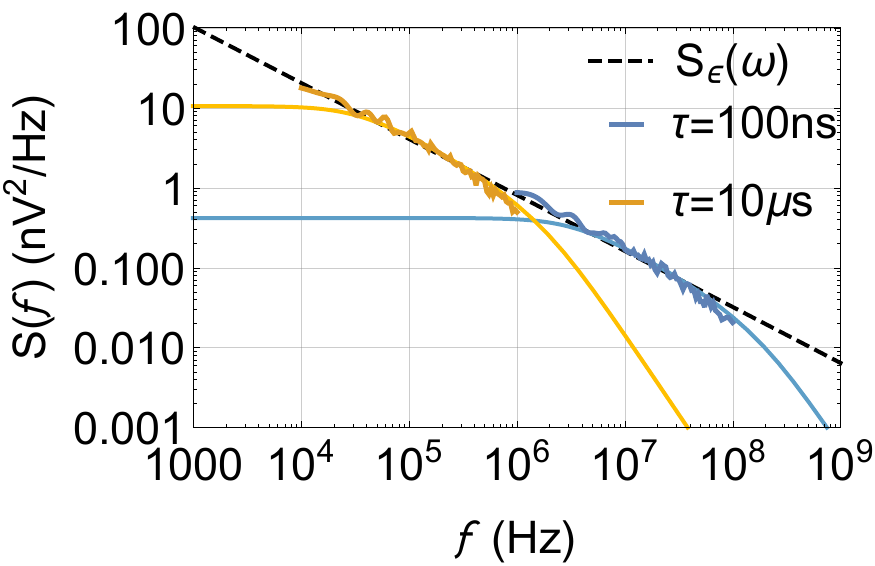}
	\caption{(Color online.) PSDs used in the simulations for two gate times, $\tau$ = 100ns and $\tau$ = 10$\mu$s. The noisy thick solid lines show the $1/\omega^{0.7}$ power spectral distributions in the band $1/10\tau$ to $10/\tau$ calculated numerically from a set of 100 temporal noise traces. For clarity, outside that frequency band we plot with thin solid lines the \emph{analytical} Lorentzian behavior of these PSDs outside the $1/f$ bandwidth (the numerically calculated PSD has a slightly shifted low-frequency plateau which is purely an artifact of the finite duration of the time traces and it also has a white noise tail at high frequency which is purely an artifact of the discretization of time in the numerical integration of the autocorrelation function). The dotted line shows $1/\omega^{0.7}$ for reference.}\label{fig:psd_plot}
\end{figure}

There are two main noise sources for singlet-triplet qubits: fluctuations in the magnetic field gradient, $\delta h_i$, and fluctuations in the detunings, $\delta\varepsilon_i$. We now simulate the evolution of the two-qubit singlet-triplet system when targeting maximally entangling gates at the interacting sweet spot. The fluctuations in $h$ are predominantly a low-frequency noise source, with its power spectral density (PSD) $S_h(\omega)\propto 1/\omega^{2.6}$ \cite{Gossard2012}. We thus model the noise in $h$ as quasistatic with a standard deviation of 8 neV \cite{Bluhm2010} (4.2 neV\cite{Eriksson2014}) in GaAs (Si). Detuning fluctuations due to charge noise also contains a quasistatic contribution, $\delta\varepsilon_i^{\mathrm{(QS)}}$, with a standard deviation of $8\mu\text{V}\times 1\text{eV}/9.4\text{V}$ (6.4$\mu$eV) in GaAs\cite{Nichol2017} (Si\cite{Eriksson2014}), but in addition includes higher frequency noise, $\delta\varepsilon_i^{\mathrm{(1/f)}}$ with a PSD $S_{\varepsilon}(\omega)\propto 1/\omega^{0.7}$ \cite{Dial2013}. We generate this ``$1/f$" noise by superimposing random telegraph noise (RTN) traces with a range of switching rates $\nu$ and varying amplitudes $(1/2\nu)^{\frac{0.7-2}{2}}$. The total PSD is then computed in order to scale the noise to the experimentally reported magnitude, $0.09~\textrm{neV}/\sqrt{\textrm{Hz}}$ ($10.04~\textrm{neV}/\sqrt{\textrm{Hz}}$) at 1 MHz for GaAs\cite{Dial2013} (Si\cite{Vavilov2017}). The resulting PSD is shown in Fig.~\ref{fig:psd_plot}. In GaAs, the noise PSD has been measured to be proportional to $1/\omega^{0.7}$ for a range 50 kHz $< \omega <$ 1 GHz\cite{Dial2013,Nichol2017}.

For ease of computation when solving for the evolution operator, we assume that the effect of the $1/f$ noise on the evolution operator of a gate of time length $\tau$ is predominantly from the frequency band ranging from $1/10\tau$ to $10/\tau$. We absorb noise at lower frequencies into the quasistatic contribution and we ignore noise at higher frequencies since we confirmed numerically that it is too fast compared to the entanglement dynamics of the evolution to have a noticeable influence, essentially averaging itself out. We then numerically construct the $1/f$ noise from ten RTNs with switching rates logarithmically distributed uniformly in this range. The generated noise is $\propto 1/\omega^{0.7}$ in the relevant bandwidth and is Lorentzian elsewhere. Fig.~\ref{fig:psd_plot} shows an example of the noise generated in this way when $\tau=100$ns and $\tau=10\mu$s.

The noisy gate is constructed by computing the time ordered exponential of Eq.~\eqref{Two-qubit_Hamiltonian} where $\varepsilon_i(t) = \varepsilon_{ss} + \delta\varepsilon_i^{\mathrm{(QS)}} + \delta\varepsilon_i^{\mathrm{1/f}}(t)$ and $h_i = h + \delta h_i$, resulting in $U'$. We then compare $U'$ to the same operation in the absence of noise $U$ using the averaged two-qubit fidelity defined in Ref.~\onlinecite{Cabrera2007}.

\begin{figure}
	\includegraphics[width = \columnwidth]{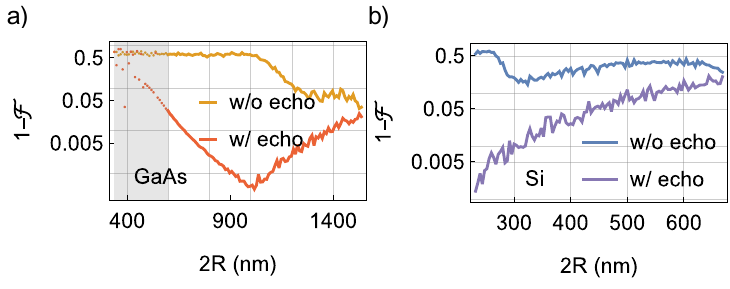}
	\caption{(Color online.) Infidelity of maximally entangling gates operated at the sweet spot, with and without an echo pulse, subject to both $\delta h$ and $\delta\varepsilon(t)$ noise for the range $R_\mathrm{min} < R < R_\mathrm{max}$ in \textbf{(a)} GaAs and \textbf{(b)} Si.}\label{fig:fig4}
\end{figure}

While the sweet spot provides protection against charge fluctuations, it does not suppress $\delta h_i$ noise at all.  However, the quasistatic nature of the magnetic field noise allows its effects to be suppressed with standard echo techniques. For example, a $\pi$-pulse about $y$ applied to both qubits halfway through the maximally entangling gate operation suppresses (though not completely removes) the errors due to both $\delta h$ and the DC component of $\delta J$ since it anticommutes with the dominant error terms produced by those noises, yet preserves the entanglement because it commutes with the non-local interaction. We compute gate fidelities both with and without the $\pi$-pulses at the halfway point, assuming errorless single qubit gates in our simulations, a reasonable approximation since single-qubit fidelities near 99\% are accessible\cite{Nichol2017}.  The results are shown in Figs.~\ref{fig:fig4}a and \ref{fig:fig4}b.

For our GaAs parameters, the magnetic field noise is much stronger than the residual charge noise coupled in through higher order derivatives at the sweet spot, so unless a pulse sequence is used to suppress magnetic noise, the optimal strategy is simply to use a sweet spot at the largest possible $R$, resulting in the shortest gate times, so as to reduce the accumulation of magnetic noise. However, if a simple echo is used to reduce the magnetic noise its residual effects become comparable to the residual charge noise. Reducing the interqubit distance decreases higher order sensitivity to charge perturbations at the cost of increasing the time over which the residual magnetic field noise accumulates, a tradeoff resulting in an optimal distance for a sweet spot operation of about 1 $\mu$m, corresponding to a gate time of about 950ns, and yielding a fidelity of 99.96\%. For our Si parameters, due to the generally faster gate times and larger charge noise, the two types of error are already comparable without using a pulse sequence, and the tradeoff results in an optimal distance of $2R\approx$ 324nm, corresponding to $\tau\approx$ 830ns and a fidelity near 86\%. Once a pulse sequence is used to further suppress magnetic noise, residual charge noise dominates and the optimal strategy is simply to use the sweet spot at the smallest possible $R$, resulting in the smallest second derivatives. Then one obtains fidelities of up to 99.86\%.

\section{Conclusion}
Using the Hund-Mulliken model for two capacitively coupled singlet-triplet qubits, we have found a symmetric outward detuning, called the interacting sweet spot, where the effective exchange is insensitive to charge noise. This is particularly useful for combatting high-frequency noise that is difficult to correct with existing standard pulse sequences. We simulate the evolution at the sweet spot under realistic charge and magnetic field noise and show maximally entangling gates above the current 90\% fidelity\cite{Nichol2017} are accessible. By using this interacting sweet spot to perform two-qubit gates and a noninteracting sweet spot \cite{Reed2016, Martins2016} to perform the single-qubit gates, one can ensure that, with the exception of the ramping times, the qubits remain protected against charge noise for the entire duration of the computation.

\section*{Acknowledgments}
This material is based upon work supported by the National Science Foundation under Grant No.~1620740, and addition of the Si results was supported by the Army Research Office (ARO) under Grant Number W911NF-17-1-0287. MAW acknowledges support from the UMBC Office of Undergraduate Education through an Undergraduate Research Award.

\bibliographystyle{apsrev4-1}
\bibliography{library2}

\end{document}